\documentclass[12pt,preprint]{aastex}

\usepackage{epsfig}

\shorttitle{Tracing the development of dust}
\shortauthors{Lebzelter et al.}

\begin{document}

\title{Tracing the development of dust around evolved stars:
The case of 47 Tuc}

\author{Th. Lebzelter, Th. Posch}
\affil{Institut f\"ur Astronomie, T\"urkenschanzstra{\ss}e 17, A-1180 Wien,
Austria}
\email{lebzelter@astro.univie.ac.at}

\author{K. Hinkle}
\affil{NOAO, P.O.Box 26732, Tucson, AZ 85726, USA}

\author{P.R. Wood}
\affil{RSAA, Mount Stromlo Observatory, Weston, ACT 2611, Australia}

\author{J. Bouwman}
\affil{Max-Planck-Institut f\"ur Astronomie (MPIA), K\"onigstuhl 17, D-69117 Heidelberg,
Germany}

\begin{abstract}

We observed mid-infrared (7.5--22\,$\mu$m) spectra of AGB stars in
the globular cluster 47 Tuc with the {\em Spitzer}\/ telescope and
find significant dust features of various types. Comparison of the
characteristics of the dust spectra with the location of the stars
in a logP-$K$-diagram shows that dust mineralogy and position on
the AGB are related.  A 13\,$\mu$m feature is seen in spectra of
low luminosity AGB stars.  More luminous AGB stars show a broad
feature at 11.5\,$\mu$m. The spectra of the most luminous stars are
dominated by the amorphous silicate bending vibration centered at
9.7\,$\mu$m. For 47\,Tuc AGB stars, we conclude that early on the
AGB dust consisting primarily of \mbox{Mg-,} Al- and Fe oxides is
formed.  With further AGB evolution amorphous silicates become the
dominant species.

\end{abstract}


\clearpage

\section{Introduction}

Stars on the Asymptotic Giant Branch (AGB) represent the final
evolutionary stage of low- to intermediate mass stars. The large
AGB mass-loss rates play a key role in the cosmic circuit of matter.
This mass loss consists of both gas and (sub-)micron-sized solid "dust"
particles.  Dust in the circumstellar shells of AGB stars leads to
a significant change in the overall spectral energy distribution
compared to dust-free objects. It attenuates stellar radiation
in the blue and visual range and re-radiates corresponding
emission at mid-infrared wavelengths (`infrared excess'). The
spectrum of the infrared excess contains broad spectral features
characteristic of specific dust species.

Up to now, dust properties of AGB stars could be determined only
for samples in the galactic disk and bulge (e.g.\ Omont et al.\
2003).  The IRAS and ISO missions created large databases of mid-
and far-IR spectra of dust-producing stars.  Analysis of these spectra
led to basic insights e.g.\ into the relation between dust
properties and stellar C vs.\ O abundances or total mass loss rates
(e.g.\ Molster \& Waters 2003).  However, as the disk and bulge
samples are inhomogeneous with respect to individual metallicities
and masses -- which both influence the production of dust -- several
questions remained open. One main open issue concerns the interrelation
between the dust composition, stellar variability, and evolutionary
status of an AGB star.  In contrast to stellar samples in the
galactic disk and bulge, globular clusters (GCs) make possible
observation of stellar samples with well-defined, homogeneous
parameters.

\section{Sample selection and data reduction \label{s:samples}}

Among the GCs, 47 Tuc (NGC 104, [Fe/H]$=-$0.76, Briley et al. 1995)
is especially suited for a study correlating AGB stellar parameters
and dust spectral signatures. This cluster has a well populated AGB
with indications of dust formation due to considerable IR excesses
(van Loon et al.\ 2006, Ramdani \& Jorissen 2001).  Lebzelter et
al.\ (2005) and Lebzelter \& Wood (2005) were able to assign
fundamental or overtone pulsation modes to most of the 47 Tuc AGB
variables. Overtone pulsators exclusively populate the lower part
of the AGB, while fundamental mode pulsators are only found at high
luminosities. The NIR velocity amplitude was found to strongly
increase during the transition from overtone to fundamental mode
and to further increase along the fundamental mode sequence with
luminosity.  Based on these findings, we selected 10 targets
distributed over the AGB pulsation modes and luminosity range 
for observation with Spitzer.  One star, V13\footnote{Throughout
this paper, variables are named according to Clement et al (2001).},
shows a short primary and a long secondary period (Lebzelter et
al.\ 2005). To this sample we added V18, a star located between the
fundamental and the overtone mode pulsators with a strong infrared
excess (Ramdani \& Jorissen 2001). It was suggested by Lebzelter
et al.\ (2005) that this star is currently in the luminosity minimum
following a thermal pulse. The upper left panel of Fig.\ 1 gives
an overview on the location of the various variables in the logP-$K$
diagram.

For these 11 targets low resolution {\em Spitzer}\/ spectra have been obtained between $7.6-21.7$ $\mu$m during Cycle 1. Our spectra are based on the {\tt droopres} products processed through the S13.2.0 
version of the {\it Spitzer\,} data pipeline.  Partially based on 
the {\tt SMART} software package \citep{Hig04}, these intermediate 
data products were further processed using spectral extraction tools 
developed for the "Formation and Evolution of Planetary Systems" 
(FEPS) {\it Spitzer\,} science legacy team.  
The spectra were extracted using a 6.0 pixel and 5.0 pixel 
fixed-width aperture in the spatial dimension for the observations 
with the first order of the short- ($7.6-14$ $\mu$m) 
and the second and third order of the long-wavelength ($14-21.7$ $\mu$m) 
modules, respectively. The background was subtracted 
using associated pairs of imaged spectra from the two nodded positions 
along the slit, also eliminating stray light contamination and anomalous dark 
currents. Pixels flagged by the data pipeline as being "bad" were replaced with 
a value interpolated from an 8 pixel perimeter surrounding the errant 
pixel. The low-level fringing in the low-resolution spectra 
at wavelengths longer than 20~$\mu$m was removed using the 
{\tt irsfinge} package \citep{Fred03}.
The spectra are calibrated using a spectral response function derived 
from IRS spectra and Cohen stellar models for a suite of calibrators 
provided by the {\it Spitzer\,} Science Centre. 
To remove any effect of pointing offsets, we matched orders 
based on the wavelength dependend point spread function of the IRS instrument, 
correcting for possible flux losses. The relative errors between 
spectral points within one order are dominated by the 
noise on each individual point and not by the calibration.  We 
estimate a relative flux calibration across an order of 
$\approx 2$~\% and an absolute calibration error between 
orders/modules of $\approx 5$~\%.

The mid-IR (MIR) SED consists of a photospheric continuum component
and various dust features on top of it. To obtain a clear view on
the signatures of the dust we subtracted a blackbody curve representing
the photosphere.  For this purpose we obtained parallel near infrared
photometry (JHKL) at Siding Spring Observatory (SSO). The blackbody
temperatures (3400-3900\,K) were derived from $J-K$ values using
the relation by Houdashelt et al.\ (2000). The flux levels of the
blackbodies were chosen such as to fit the near infrared flux levels.
For sample stars with low mass loss rates we assume that the infrared
excess will not significantly affect $J-K$.  Fig.\,2 gives an example
of a spectrum before and after blackbody subtraction.  In the
following we will present and discuss only continuum subtracted
spectra.  The chosen continuum subtraction method naturally introduces
some uncertainty in the derived residual dust emission, but this
uncertainty does not affect the positions and relative strengths
of the various dust features on which this paper is focused.

\section{The MIR spectra of AGB stars in 47 Tuc \label{s:MIR}}

\subsection{Phenomenology of the dust spectra} \label{scenario}

V5, V6 and V7 can be fitted nicely with single blackbodies of about
3700 to 3900\,K and must be nearly dust-free.  Dust features were
detected in the remaining 8 of the 11 sample stars.  The three
dominant dust features are located at approximately 9.7, 11.5 and
13\,$\mu$m. Additionally, broad features at 18--20\,$\mu$m are
visible. However, not all features are visible in all stars. In
Fig.\ 1, we show the dust spectra and the location of the corresponding
stars in the log\,P-$K$ diagram. V13 is presented separately in
Fig.\ 2.  From the logP-$K$ values we can form three groups among
the sample stars. The spectra of pairs V5/V6, V4/V8 and V1/V2 are
very similar, suggesting a typical spectrum is associated with a
specific location in the logP-$K$-diagram.  An interesting case is
V3 with similar logP-$K$ values as V1/V2.  The V3 spectrum has a
very weak dust feature sitting on top of a broad mid-IR emission
with a wavelength dependent slope. This slope can be roughly
approximated by a blackbody curve on top of which a weak 9.7\,$\mu$m
feature stands out.  While not obvious, this is the same spectral
signature as in V1/V2.  In \ref{variabi} we further discuss these
differences as a function of stellar variability.

The change of the dust features along the AGB is in agreement with
earlier studies on the relation between dust properties and stellar
parameters.  While previous works on this subject (e.g.\ Onaka et
al.\ 1989, Sloan \& Price 1996, Hron et al.\ 1997, Heras \& Hony
2005) refer to an inhomogeneous samples of stars, our GC spectra
minimize possible effects of such inhomogeneities.  At the lower
luminosity end of the AGB, hardly any dust is present.  Effective
dust production seems to start only after the star reaches the
luminosity of V21 corresponding to a value of approximately
2000\,L$_{\sun}$ (cf.\ Lebzelter \& Wood 2005). The 13\,$\mu$m
feature clearly dominates the spectrum at this luminosity.  Changing
from V21 to V4 brings about a switch in pulsation mode from overtone
to fundamental mode and a strong increase in the observed velocity
amplitude. The related change in the MIR spectrum is an increase
in the 11.5\,$\mu$m feature, a decrease in the strength of the
13\,$\mu$m, and the occurrence of a 9.7\,$\mu$m feature. Going to
even higher luminosities leads to V1 with its dominating 9.7\,$\mu$m
feature and only a very weak indication for the 11.5\,$\mu$m feature,
sitting on the long wavelength side of the 9.7\,$\mu$m emission.

Summarizing these observations, we can say that the 13\,$\mu$m
feature appears in the least evolved stars, later looses importance
relative to the 11.5\,$\mu$m peak and a feature at 9.7\,$\mu$m and
finally vanishes completely in the highest luminosity stars. The
latter are dominated completely by the 9.7\,$\mu$m emission.  

\subsection{Identification of the dust features}

Among the observed emission bands only the 9.7\,$\mu$m feature has
a generally accepted identification, amorphous Mg-Fe-silicate (see,
e.g., Dorschner et al.\ 1995 for optical constants).  No consensus
exists on the carriers for the broad 11.5\,$\mu$m band, the narrow
13\,$\mu$m band, and the various features between 17 and 20\,$\mu$m.
Since the sample stars are all spectral type K or M the dust species
must be oxides.

First detected in IRAS spectra (Little-Marenin \& Little 1988),
the 13\,$\mu$m band was identified as part of a `three-component-feature'
consisting of a 9.7, $\sim$11, and the 13\,$\mu$m band itself.  Our
spectra of 47 Tuc V4 and V8 show the `three-component-feature'.
The 13\,$\mu$m feature is likely an Al-O stretching vibration.  The
carrier has been identified as either corundum ($\alpha$-Al$_2$O$_3$,
Glaccum 1995, DePew, Speck \& Dijkstra 2006) or spinel (MgAl$_2$O$_4$,
Posch et al.\ 1999, Fabian et al.\ 2001, Heras \& Hony 2005).

The origin of the broad 11.5\,$\mu$m band in spectra of O-rich AGB
stars has been extensively discussed (e.g.\ Stencel et al.\ 1990,
Lorenz-Martins \& Pompeia 2000, Egan \& Sloan 2001, Posch et al.\
2002, Heras \& Hony 2005). Although it is especially difficult to
identify uniquely such a broad solid state band, there remains
little doubt that amorphous Al$_2$O$_3$ is the band carrier in this
case.\footnote{ Amorphous Al$_2$O$_3$ is sometimes called `corundum'
in the literature.  However, this name should be reserved for the
crystalline $\alpha$-phase of Al$_2$O$_3$.}

The band (or double-band) at 18--20\,$\mu$m is most distinctly seen
in V1 (18\,$\mu$m), V4 (18+20\,$\mu$m), V8 (20\,$\mu$m), V13
(20\,$\mu$m), V18 (18+20\,$\mu$m) and V21 (20\,$\mu$m). The observed
differences in the positions of the feature's maxima probably
indicate a transition from a crystalline oxide (with a composition
between MgO to FeO, see Posch et al. 2002) as the dominant opacity
source in the 19-21\,$\mu$m region to an amorphous silicate's bending
vibration as the primary emission mechanism.  Sloan et al. (2003)
found a correlation between the strengths of the 13 and 20\,$\mu$m
features.  

In Fig.\ 2 we show the spectrum of V13 with the band identifications
for the oxide dust features suggested above.  The spectrum of V13
is further discussed in section \ref{v13v18}.

\subsection{Variability of the mid-IR spectra} \label{variabi}

At least for fundamental-mode AGB pulsators cyclic variability leads
to significant changes in size and T$_{\rm eff}$ of the star.  These
could effect the observed MIR spectra through cyclic changes in
both dust production and photospheric flux.  From Spitzer we have
only one observational epoch for each star.  However, additional
information provides indications of the impact on the mid-infrared
spectrum.

Our sample contains the three miras, V1, V2, and V3, all sharing a
similar location in the logP-K-diagram. Their spectra show the same
dust features, but with a very different contrast against the
photospheric background.  From the simultaneous $K$ photometry, we
estimate that V1 was observed close to light maximum, while V2 and
V3 were close to minima.  We speculate that the background is related
to the pulsation phase.  This is seen in model atmospheres (Aringer
2005).  Phase dependent background can possibly explain the difference
between our spectrum of V8 and one published by van Loon et
al.\,(2006). While dust features are quite obvious in the Spitzer
observation, the van Loon spectrum taken from the ground shows no
features at all. The $K$ amplitude of V8 is too small to safely
estimate the phase from the observed $K$ magnitude.

\subsection{Two peculiar cases: V13 and V18} \label{v13v18}

Two stars do not seem to fit in the dust evolution scenario described
in \ref{scenario}: V13 and V18 which both have IR excess emission
much too strong in relation to their position in the logP-K diagram.
The residual dust emission spectrum of V13 (Fig.\ 2) is almost
devoid of a 10\,$\mu$m band. This alone is not surprising, because
it also is the case for V21. However, V13 is almost 1\,mag fainter
in $K$ than V21. The spectra of both V13 and V21 are characterized
by pronounced 11.5, 13 and 20\,$\mu$m peaks.  V18 shows a spectrum
very similar to the mira V1, i.e.~it is dominated by a 9.7$\mu$m
silicate dust feature, yet is $\sim$1.25 magnitude fainter at $K$.

What properties of V13 enable the formation of dust?  The
only remarkable pulsational characteristic of V13 is its long
secondary period.  This long period is also clearly visible in the
radial velocity (Lebzelter et al.\ 2005) as well as photometric
variation.  The length of the secondary period has not be determined
accurately but seems to fall onto the Wood et al.\ (1999) P-L
sequence "D".  The interpretation of this sequence in terms of
stellar pulsation or binarity is still not clear (e.g.\ Wood 2006).

We interpret the dust spectrum of V18 in the light of the earlier
suggestion that this star is currently undergoing a thermal pulse
event.  V18 would then have an interpulse luminosity and dust
spectrum similar to that of an 47 Tuc mira, e.g. V1.  According to
evolution models of thermally pulsing AGB stars by Vassiliadis \&
Wood (1993), a significant disturbance of the typical pulsation
period of a star occurs during the thermal pulse (cf.\ their Fig.\
3), which would explain the strange location of V18 in the logP-$K$
diagram.

\section{Consequences for the evolution of dust on the AGB \label{s:Cons}}

Our evolutionary sequence of the dust spectra indicates that
oxygen-rich AGB stars first form dust rich in oxides of aluminum
and magnesium (specifically, \mbox{Al$_{2}$O$_{3}$-}, \mbox{MgAl$_{2}$O$_{4}$-}
and (Mg,Fe)O-). As a star climbs up the AGB, the original
aluminum-magnesium oxide dust contributes less to the emergent dust
shell spectra, while the relative strength of amorphous silicate
bands, especially the one at 9.7\,$\mu$m, increases. This scenario
is in agreement with considerations published by Stencel et al.\
(1990) linking this evolutionary sequence to the higher electron
affinity of Al (compared to Si) to oxygen. It is the high binding
energy of solid Al oxides which enables them to condense at much
higher temperatures than silicates (see, e.g., Gail 2003, Fig.\ 4).
Once the Al oxides have formed, silicates will start to precipitate
on their surfaces which gradually leads to a transition to a silicate
dominated dust spectrum.  Hron et al.\ (1997) pointed out that the
13\,$\mu$m feature reaches significant strength only in a narrow
range of effective temperatures and optical depths.  They proposed
an evolutionary sequence in the course of which the carrier of this
band would be gradually incorporated into amorphous silicates with
increasing mass-loss rate. Most recent model calculations by Woitke
(A\&A submitted) indicate that Al$_{2}$O$_{3}$ can exist closer to
the star while silicates are formed at larger distances, i.e. once
effective mass loss has set in.

A similar scenario for aluminum-magnesium oxide/silicate dust evolution
has been proposed by Onaka et al.\ (1989). They derive a relation
between the asymmetry factor of the light curve and the importance
of the various dust features with the stars exhibiting the strongest
asymmetries showing the most pronounced silicate emission. The
asymmetry of the light curve can be related to the strengths of the
shock front occurring during the cyclic pulsation as discussed by
Onaka and coworkers. Lebzelter et al.\ (2005) showed that the near-IR
velocity amplitude increases along the AGB of 47 Tuc.  Thus a
relation between pulsational properties and the dust composition
can be expected.  The necessity of taking into account pulsation
characteristics as an additional factor beside luminosity is
illustrated by comparing the spectra of V21 and V4 (Fig.\ 1). The
overall shape of the mid-infrared spectrum -- but not the dust
features that occur -- seems additionally to depend on the pulsational
phase.  A more complete discussion of the individual dust spectra
will be given in a forthcoming paper, where we will give a more
detailed presentation of the spectra and a comparison with combined
atmospheric and dust models.

\begin{small}
{\acknowledgments
This work is based on observations made with the Spitzer Space
Telescope, which is operated by the Jet Propulsion Laboratory,
California Institute of Technology under a contract with NASA.
Support for this work was provided by NASA through an award issued
by JPL/Caltech.  TL acknowledges funding through FWF project P18171.
PRW received partial funding support from an Australian Research
Council Discovery Grant. JB acknowledges support from the
EU Human Potential Network contract No. HPRN-CT-2002000308.
We thank H.-P.\ Gail and J.\ Hron for
fruitful discussions on this paper.}
\end{small}

\clearpage

\begin{figure}
\epsscale{0.95}
\plotone{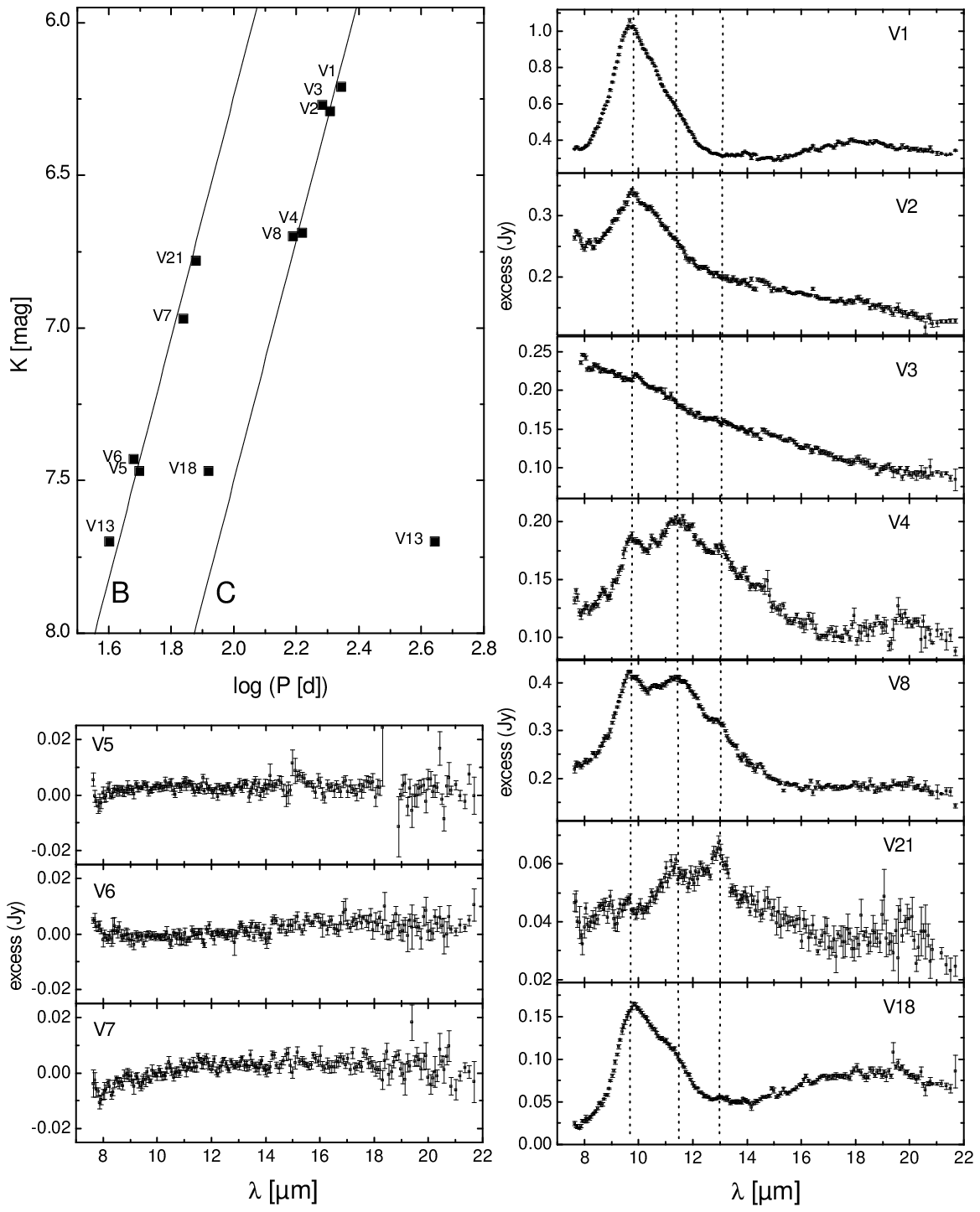}
\caption{Development of dust spectra in 47 Tuc.
{\it Right and lower left:} 
Spitzer spectra of the AGB stars in 47 Tuc (for V13 see Fig.\ 2)
with blackbody curves 
subtracted (see text). Spectra are plotted as individual data
points with errors bars derived from the difference between the
two nod positions. The features at
9.7, 11.5 and 13\,$\mu$m
are indicated by vertical dashed lines. {\it Top left corner:} logP-K-diagram
of the sample stars adapted from Lebzelter et al.\ (2005). Solid
lines mark the locations of overtone (B) and
fundamental mode (C) radial pulsators (Wood et al.\ 1999). V13 is plotted with
both of its periods. 
   \label{f:ov}}
\end{figure}

\begin{figure}
\epsscale{0.85}
\plotone{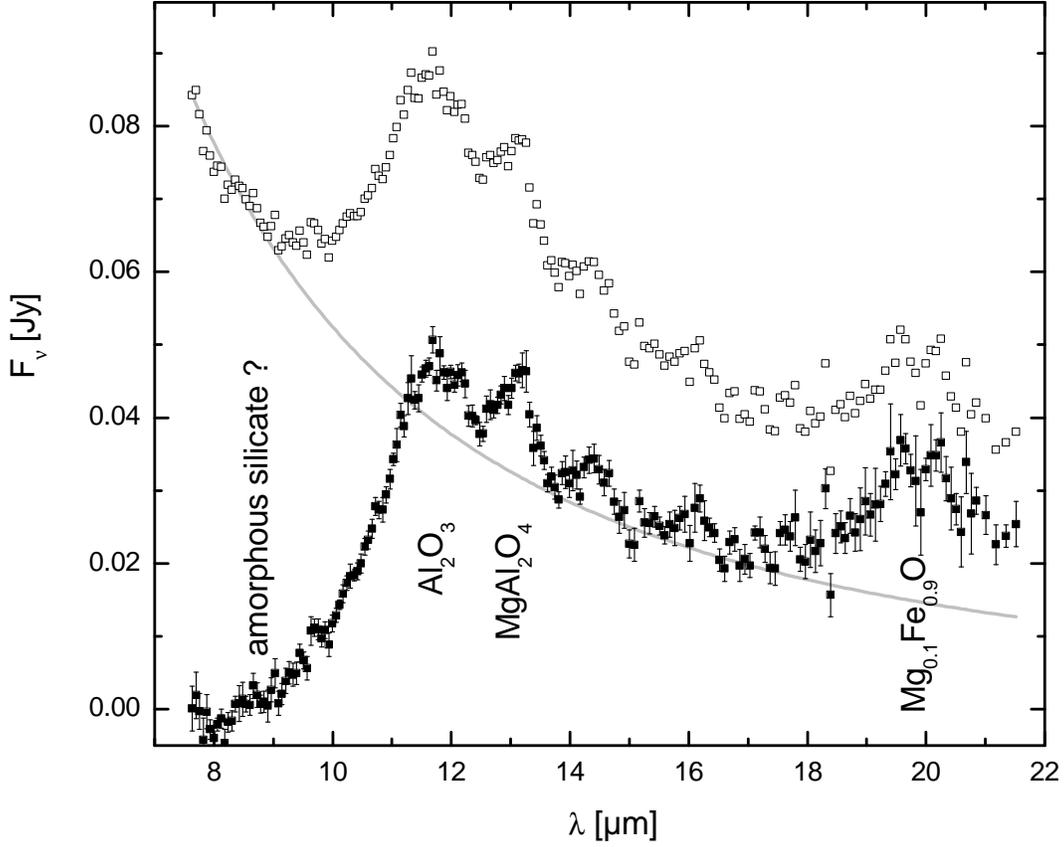}
\caption{The Spitzer mid-IR spectrum of V13 (open symbols) and the
residual dust emission (filled symbols). The grey line gives the
blackbody used for continuum subtraction. Error bars are derived
from the difference of the two nod positions. The indicated dust
feature peak positions are from the optical constants published
for amorphous Al$_2$O$_3$ (Eriksson et al.\ 1981), 
crystalline MgAl$_2$O$_4$ (Fabian et al.\ 2001),  
Mg$_{0.1}$Fe$_{0.9}$O (Henning et al.\ 1995), and 
(hardly detectable) amorphous silicate (Ossenkopf et al.\ 1992). 
\label{f:v13}}
\end{figure}

\end{document}